\documentclass[aps,prd,superscriptaddress,nofootinbib,preprintnumbers]{revtex4}
\usepackage{amsmath}
\usepackage{graphicx}
\usepackage{color} 
\usepackage{amsmath}
\newcommand{\lesssim}{\mathrel{\mathpalette\vereq<}}

\newcommand{\chushi}[1]{}
\newcommand{\1}{\mbox{1}\hspace{-0.25em}\mbox{l}}

\begin{document}
\title{{\bf  Proving Rho Meson be a Dynamical Gauge Boson of Hidden Local Symmetry
}
 \vspace{5mm}}

\author{{Koichi Yamawaki}} \thanks{
      {\tt yamawaki@kmi.nagoya-u.ac.jp}}
      \affiliation{ Kobayashi-Maskawa Institute for the Origin of Particles and 
the Universe (KMI), Nagoya University, Nagoya 464-8602, Japan.}

\begin{abstract} 
The rho meson  has long been successfully 
identified with a dynamical gauge boson of the Hidden Local Symmetry (HLS)  $H_{\rm local}$ in the nonlinear sigma model $G/H$ 
gauge equivalent to the model having the symmetry 
$G_{\rm global}\times H_{\rm local}$, with $G= [SU(2)_L \times SU(2)_R]\simeq O(4), H=SU(2)_{V}\simeq O(3)$,  however under a hitherto unproven assumption that its kinetic term is dynamically generated, together with an ad hoc  choice of the auxiliary field parameter ``$a=2$''.  We prove this assumption, thereby solving the long-standing mystery:  The rho meson kinetic term is generated simply by  the large $N$ limit  of the Grassmannian model $G/H=O(N)/[O(N-3)\times O(3)] $ gauge equivalent to 
$O(N)_{\rm global}\times [O(N-3)\times O(3)]_{\rm local}$, extrapolated to $N=4$,   $O(4)_{\rm global}\times O(3)_{\rm local}$, with all the phenomenologically successful ``$a=2$ results'' i.e.,  $\rho$-universality, KSRF relation and  the Vector Meson Dominance,  
realized  {\it independently of the parameter ``$a$''.}  This in turn  establishes validity of the  large $N$ dynamics at quantitative level directly by the experiments. The relevant cutoff reads $\Lambda \simeq 4 \pi F_\pi$ for $N=4$, which is regarded  as a matching scale of the HLS as a ``magnetic dual''  to QCD. Skyrmion is stabilized by  such a  dynamically generated rho meson without recourse to the underlying QCD, further signal of the duality. The unbroken phase with massless rho meson may be realized as  a novel chiral restored hadronic phase in hot/dense QCD.
\end{abstract} 
\maketitle

\section{Introduction}
Since its proposal  \cite{Bando:1984ej,Bando:1985rf,Bando:1984pw,Fujiwara:1984mp,Bando:1987ym}  
(for reviews see Ref.\cite{Bando:1987br,Harada:2003jx,Yamawaki:2016qux}),  identifying the rho meson 
as a dynamical gauge boson of the Hidden Local Symmetry (HLS) $H_{\rm local}$ has been widely accepted in the model  
having the symmetry  $G_{\rm global}\times H_{\rm local}$,
with $G=[SU(2)_L\times SU(2)_R]\simeq O(4)$ and $H=SU(2)_V\simeq O(3)$, where its Lagrangian consists of two independent invariants  ${\cal L}_{\rm HLS}={\cal L}_A + a {\cal L}_V$, with $a$ arbitrary parameter. This is gauge equivalent to the nonlinear sigma model, ${\cal L}_{\rm CCWZ}$, $\grave{a}$ la Callan-Coleman-Wess-Zumino (CCWZ)~\cite{Coleman:1969sm,Callan:1969sn}  based on the manifold $G/H$: 
In the absence of the kinetic term of the HLS gauge boson it is merely an auxiliary field such that 
${\cal L}_V=0$, and ${\cal L}_{\rm HLS}={\cal L}_A={\cal L}_{\rm CCWZ}$ after gauge fixing.  
 
Once we {\it assume}, however, that its kinetic term, ${\cal L}_{\rm kinetic}$, is generated at quantum level by the dynamics of the nonlinear sigma model itself,  thereby put by hand to the Lagrangian, ${\cal L}_{\rm HLS}\Rightarrow {\cal L}_A + a {\cal L}_V + {\cal L}_{\rm kinetic}$, novel physics come out~\cite{Bando:1984ej,Bando:1985rf,Bando:1984pw,Fujiwara:1984mp,Bando:1987ym, Bando:1987br,Harada:2003jx,Yamawaki:2016qux}: All the successful phenomenological results, such as the universality of the rho meson coupling ($\rho$ universality), the Kawarabayashi-Suzuki-Ryazzudin-Fayyazudin (KSRF) relation, the Vector Meson Dominance (VMD), are derived for a {\it particular parameter choice $a=2$}  in the  resultant Lagrangian ({\it at tree level}),  in such a way that $ a {\cal L}_{ V}$ becomes the HLS gauge-invariant mass terms of the $\rho$ meson which contains $\rho$ mass, $\rho$ couplings, additional $\pi$ self-couplings, etc..

For all the phenomenological success of the HLS model of the rho meson, however, the basic assumption of the dynamical origin of the kinetic term and the particular parameter choice $a=2$ has never been proved within the dynamics of HLS model itself.~\footnote{
Alternatively, we may simply assume that the kinetic term is already generated by the underlying theory, QCD in the case at hand, regarding the HLS model as a low energy effective theory, including loop corrections in the sense of the derivative expansion~\cite{Harada:2003jx}.
In this case, however,  the parameter $a$ (renormalized one) is a completely free parameter to be adjusted to $a=2$ by hand within the HLS model framework. 
It can be determined  only with additional information, by ``Wilsonian matching'' with the QCD parameters at UV scale $\Lambda$ as input, $a(\Lambda^2)\sim 1$, which then predicts $a(\mu^2=M_\pi^2=0)=2$  as the  infrared value (massless $\pi$ on-shell)  through the (one-loop) renormalization group for $F_\pi^2(\mu^2=0)$ due to $\pi$ loop alone  in  the HLS model itself (see Re.~\cite{Harada:2003jx}
and references cited therein). 
}

In this paper  
\footnote{
Preliminary results of this paper were given  as the
supplementary ones in Ref.\cite{Yamawaki:2018jvy} which is mainly addressed to a  subject on a possible dynamical gauge boson of HLS within the Higgs sector of the Standard Model,  different from the present one, that in the QCD, but the details of the relevant calculations in the present paper may be found in  Ref.\cite{Yamawaki:2018jvy}. }
we resolve this long-standing mystery  as simply a {\it consequence of the nonperturbative dynamics of the large $N$ limit of the nonlinear sigma model} based on the Grassmannian manifold $G/H=$$O(N)/[O(N-p)\times O(p)$, with $p=3=$ fixed, which is reduced to the relevant case $G/H=O(4)/O(3)\simeq SU(2)_L\times SU(2)_R/SU(2)_V$ for the extrapolation $N\rightarrow 4$, with the rho meson being the dynamical gauge boson of $O(3)_{\rm local}\simeq [SU(2)_V]_{\rm local}$.

It is in fact well known that the HLS gauge bosons in many nonlinear sigma models, such as the $CP^{N-1}$ model with $G/H=U(N)/[U(N-1)\times U(1)]$$\simeq SU(N)/[SU(N-1)\times U(1)]$ gauge equivalent to the model $SU(N)_{\rm global} \times U(1)_{\rm local}$, do acquire kinetic term for $U(1)_{\rm local}$ gauge boson  at quantum level  in the  large $N$ limit~\cite{Eichenherr:1978qa,Golo:1978de,DAdda:1978vbw,DAdda:1978dle,Witten:1978bc,Arefeva:1980ms,Haber:1980uy,Kugo:1985jc,Bando:1987br,Weinberg:1997rv,Harada:2003jx}.
It was further  shown (in the context irrelevant to the rho meson physics, though) that the HLS gauge bosons $O(p)_{\rm local}$ and $U(p)_{\rm local}$ in the Grassmannian models $G/H=O(N)/[O(N-p)$$\times O(p)$~\cite{Brezin:1980ms}   and $G/H=$$U(N)/[U(N-p)\times U(p)]$, respectively~\cite{Brezin:1980ms,Bando:1996pg}  are dynamically generated in the large $N$ limit. 
However it was shown only in a specific parameterization ``covariant derivative'' type which is just {\it a particular $a=2$ choice} from the onset (see~\cite{Yamawaki:2018jvy} and Eq.(\ref{a-depaction})). 

Here we show that {\it not just the generation of the $O(p)_{\rm local}$ gauge boson $\rho_\mu$ but also  all the successful ``$a=2$ results''  are  direct consequences of the pure dynamics at quantum level of the large $N$ limit}  for {\it arbitrary value} of $a$, thereby resolving the long-standing mystery of the rho meson simply on the firm dynamical base. This in turn provides yet another experimental verification 
of the large $N$ reliability, this time even quantitatively, not just qualitatively~\footnote{
It is known that the large $N$ results remain qualitatively true even for the smallest  value  $N=2$ in the $CP^{N-1}$ model,  as checked by  the equivalent  $O(3)$ model exactly solvable in 2 dimensions.~\cite{Witten:1978bc}.
\label{largeN}
}

\section{Grassmaniann $N$ Extension}

   Let us define the generic HLS base  \cite{Bando:1985rf, Bando:1987br} of an  $N\times N$ real matrix field $\xi(x)=\xi({\check \rho}^{(p)})\cdot \xi({\check \rho}^{(N-p)}) \cdot \xi(\pi)$, which transforms under $G_{\rm global}\times H_{\rm local}=O(N)_{\rm global}\times [O(N-p)\times O(p)]_{\rm local}$ as $\xi(x) \rightarrow h(x) \cdot \xi(x) \cdot g^{-1}$ with $h(x) \in [O(N-p)\times O(p)]_{\rm local}\,, g \in O(N)_{\rm global}$, where $\xi(\pi)= e^{i\pi_a(x)X_a/f_\pi}$ is the  CCWZ base for $G/H=O(N)/[O(N-p)\times O(p)]$, 
with $f_\pi$ being the ({\it bare/tree-level}) decay constant of the NG boson $\pi$, while $\xi({\check \rho}^{(p)})=e^{i {\check \rho}^{(p)}(x)/f_\rho^{(p)}} $ and $\xi({\check \rho}^{(N-p)})=e^{i {\check \rho}^{(N-p)}(x)/f_\rho^{(N-p)} }$, with $ {\check \rho}^{(p)}= {\check \rho}^{(p)}_a S_a^{(p)}$ and $ {\check \rho}^{(N-p)} ={\check \rho}^{(N-p)}_a S_a^{(N-p)}$ being the would-be NG bosons to be absorbed into the HLS gauge bosons  $\rho^{(p)}_\mu$ and $\rho^{(N-p)}_\mu$ of $O(p)_{\rm local}$ and $O(N-p)_{\rm local}$, with the ({\it bare/tree-leve}) decay constant $f_\rho^{(p)}$ and $f_\rho^{(N-p)}$, respectively:
\begin{eqnarray}
\xi(x)=\xi(\pi)\, \quad \left({{\rm unitary}\,\, {\rm gauge} ,\,\,    {\check \rho}^{(N-p)}(x)= \check \rho}^{(p)}(x)=0\right).
\end {eqnarray}
Here the generators read $X_a \in {\cal G}-{\cal H}$, $S_a^{(p)} \in {\cal H}^{(p)}={\cal O}(p)$, $S_a^{(N-p)}\in {\cal H}^{{N-p}}={\cal O}(N-p)$, with ${\rm tr} (T_a T_b)= 2 \delta_{ab}$, ${\rm tr} (S_a X_b)=0$,  $T_a=\{ S_a, X_a\}=-T_a^t$. 
 
 To study the large $N$ limit in the Grassmannian models including the $CP^{N-1}$ model it is customary to  parameterize the HLS base as:\footnote{
$p\times N$ degrees of freedom of $\phi_{i\beta}$ consist of $p\times (N-p)$ of $\pi$, $p\times (p-1)/2$ of $\check \rho$, and  $p\times (p+1)/2$ of the constraints.   
}        
  \begin{eqnarray}
  \xi (x)_{\alpha \beta} 
&=&\frac{G}{N}\left(\begin{array}{c}
   \phi_{i,\beta}(x)\\
    \Phi_{k,\beta}(x)
   \end{array}
   \right), \,\, \alpha=(i,k), \,\,\beta=(j,l),\,
  i,j = 
  1,\cdots, p\,\,\, ; k,l = p+1,\cdots N \,,\\
    \xi^t\cdot \xi &=&\frac{G}{N} \left(\phi^t \phi + \Phi^t \Phi\right) 
    =\1\,, \quad \frac{G}{N} \equiv \ \frac{1}{{f_\pi}^2}, \nonumber\\
  \xi\cdot \xi^t &=& \frac{G}{N} \left(\begin{array}{cc}
   \phi \phi^t &\phi \Phi^t\\
  \Phi \phi^t& \Phi \Phi^t\end{array}\right)   
  =\left(\begin{array}{cc}
  \1_{p\times p} & {\huge 0} \\
  {\huge 0}& \1_{(N-p)\times (N-p)}
  \end{array}\right) =\1,
    \label{ONpNp}
     \end{eqnarray}
where $G\equiv N/{f_\pi}^2$ is the (bare) coupling constant to be fixed in the large $N$ limit (s.t. ${f_\pi}^2={\cal O}(N)$).

   The covariantized Maurer-Cartan one-form reads: 
 \begin{eqnarray}
  \hat  \alpha_\mu&\equiv& \frac{1}{i} D_\mu \xi  \cdot \xi^t =\frac{G}{i N} 
 \left( \begin{array}{c}
 \partial_\mu \phi- i  \rho^{(p)}_\mu \phi\\
   \partial_\mu \Phi -i  \rho^{(N-p)}_\mu \Phi
   \end{array}  
    \right) \cdot \left(\phi^t\,\, \Phi^t\right) =   {\hat \alpha}_{\mu,\perp} +  {\hat  \alpha}_{\mu,||},
\end{eqnarray}
where
 ${\hat \alpha}_{\mu,\perp}\equiv \frac{1}{2}{\rm tr} \left({\hat \alpha}_{\mu} X^a\right) X^a$, ${\hat  \alpha}_{\mu,||} \equiv \frac{1}{2}{\rm tr} \left({\hat \alpha}_{\mu} S^a\right) S^a$ are 
 \begin{eqnarray}  
  {\hat \alpha}_{\mu,\perp} = \alpha_{\mu,\perp}  =  
       \left(\begin{array}{cc}
 0&  \frac{G}{i N}  \partial_\mu \phi  \cdot \Phi^t\\
   \frac{G}{i N}  \partial_\mu \Phi  \cdot \phi^t & 0\end{array}
  \right), \quad 
  {\hat  \alpha}_{\mu,||}  = 
   \left(\begin{array}{cc}
 \frac{G}{i N}  \partial_\mu \phi \cdot \phi^t - 
 \rho^{(p)}_\mu & 0\\
  0& \frac{G}{i N}  \partial_\mu \Phi \cdot \Phi^t - 
  \rho^{(N-p)}_\mu 
  \end{array}
  \right) ,  \nonumber
    \end{eqnarray}
all  transforming homogeneously   as
 \begin{eqnarray}
 \left({\hat \alpha}_{\mu,\perp}, {\hat \alpha}_{\mu,||}  \right)  &\rightarrow& h(x) \cdot  
 \left({\hat \alpha}_{\mu,\perp}, {\hat \alpha}_{\mu,||}  \right)  \cdot h^{-1}(x)\,,
    \end{eqnarray}
  with $h(x) \in {\cal H}$ for $H =\left[O(N-p) \times   O(p)\right]_{\rm local}$.      
    
   Thus the HLS Lagrangian consists of  three independent invariants at the lowest derivative:~\cite{Yamawaki:2018jvy}  
   \begin{eqnarray}
  {\cal L}^{(N,p)}
  &=& {\cal L}_A + a^{(p)} {\cal L}_V^{(p)}+  a^{(N-p)} {\cal L}_V^{(N-p)} \,,
\label{Model}
\end{eqnarray}
where
\begin{eqnarray}
 {\cal L}_A&=&\frac{{f_\pi}^2}{4} {\rm tr} \left(
     {\hat \alpha}^2_{\mu,\perp} \right)
     =- \frac{G}{2 N}   {\rm tr} \left(\phi^t \partial_\mu \phi \cdot \Phi^t \partial^\mu \Phi  
     \right) \nonumber\\
    &=&
    \frac{1}{2}  {\rm tr}
    \left(\partial_\mu \phi   \partial^\mu \phi^t  
     +\frac{G}{N}\left(\phi \partial_\mu  \phi^t \right)^2   \right) 
        \nonumber\\
    ({\rm unitary\,  gauge})   &\longrightarrow&     {\cal L}_{\rm CCWZ}= \frac{{f_\pi}^2}{4} {\rm tr} \left({\alpha}^2_{\mu,\perp}(\pi) \right)^2
    = \frac{1}{2} \left(\partial_\mu \pi_a\right)^2  +\cdots 
    \,, \label{gaugefixedLA}
   \end{eqnarray}
with ({\rm unitary-gauge}) ${\alpha}_{\mu,\perp} \rightarrow 
 {\alpha}_{\mu,\perp}(\pi) =\partial_\mu \xi(\pi)\cdot \xi^t(\pi)=\partial_\mu \xi(\pi)\cdot \xi^\dagger(\pi) $, and
\begin{eqnarray}
 a^{(p)} {\cal L}_V^{(p)}&=&\frac{a^{(p)} {f_\pi}^2}{4}{\rm tr}
 \left(
     \left[
     {\hat \alpha}^{(p)}_{\mu,||}
     \right]^2 
     \right)=
\frac{1}{2}{\rm tr} \left[\frac{a^{(p)}}{2}\cdot \frac{N}{G}   \left(\rho^{(p)}_\mu- i \frac{G}{N} \phi \partial_\mu \phi^t 
  \right)^2\right]
    \nonumber \\
 &=& 
  \frac{(f_\rho^{(p)})^2}{4} {\rm tr}
  \left[\rho^{(p)}_\mu - \frac{\partial_\mu {\check \rho}^{(p)}}{f_\rho^{(p)} }  
- 
 \frac{i
\left[
 \partial_\mu \pi,\pi\right]}{2{f_\pi}^2} +\cdots
  \right]^2, 
   \label{LVp} 
   \end{eqnarray}
  where we should impose a bare/tree relation between the two decay constants, 
  \begin{eqnarray}
  (f_\rho^{(p)})^2= a^{(p)} {f_\pi}^2 =a^{(p)} \frac{N}{G}\,,
  \label{frho}
  \end{eqnarray}
   to normalize the kinetic term of the would-be NG boson ${\check \rho}^{(p)}$ to the canonical form, and similarly for $a^{(N-p)} {\cal L}_V^{(N-p)}$. In the unitary gauge, ${\check \rho}^{(p)}=0$, Eq.(\ref{LVp}) reads the mass term of $\rho_\mu^{(p)}$
  as usual in the HLS formalism, and so does $a^{(N-p)} {\cal L}_V^{(N-p)}$ the mass term of $\rho_\mu^{(N-p)}$. 

Here we note that in contrast to $O(p)_{\rm local}$ gauge boson, the kinetic term for the $O(N-p)_{\rm local}$ gauge boson, 
carrying index running $1,\cdots, N-p$ thus subject to all the planar diagram contributions in the large $N$ limit, is {\it not} dynamically generated~\footnote{
$O(N-p)_{\rm local}$ does not exist for  $N=4, p=3$ anyway. 
}
 and stays as an auxiliary field (i.e., ${\cal L}^{(N-p)}_V=0$) as was the case in the previous calculations  for $CP^{N-1}$ and 
Grassmannian models.
\footnote{
The $SU(N-1)_{\rm local}$ gauge boson in $CP^{N-1}$ model with $G/H=SU(N)/[SU(N-1)\times U(1)]$, which
carries the index running through $1,\cdots, N-1$, is not dynamically generated in the large $N$ limit, in contrast to the $U(1)_{\rm local}$ part~\cite{Eichenherr:1978qa,Golo:1978de,DAdda:1978vbw,DAdda:1978dle,Witten:1978bc,Arefeva:1980ms,Haber:1980uy,Kugo:1985jc,Bando:1987br,Weinberg:1997rv,Harada:2003jx}.  
The same is true for $G/H=O(N)/[O(N-p)\times O(p)]$ and
$G/H=U(N)/[U(N-p)\times U(p)]$, with  $O(N-p)_{\rm local}$ and $U(N-p)_{\rm local}$, respectively
~\cite{Brezin:1980ms,Bando:1996pg}.
 Similarly, a popular $N$ extension $G/H=O(N)/O(N-1)$ gauge equivalent to the model 
 $O(N)_{\rm global} \times O(N-1)_{\rm local}$ has no dynamical gauge boson for $O(N-1)_{\rm local}$ and is irrelevant to the rho meson.
} 
 
 Then without loss of generality the starting Lagrangian, Eqs.(\ref{Model})-(\ref{LVp}), is simplified as  
 $a^{(p)} {\cal L}^{(p)}_V$ $\equiv  a {\cal L}_V $, 
 $a^{(N-p)} {\cal L}^{(N-p)}_V=0$, $\rho_\mu\equiv \rho^{(p)}_\mu$, ${\check \rho}={\check \rho}^{(p)}$, $f_\rho^2$$\equiv (f_\rho^{(p)})^2=a f_\pi^2$, etc.:
   \begin{eqnarray}
  {\cal L}= {\cal L}_A+ a {\cal L}_V=
\frac{1}{2} {\rm tr}
 \left[ 
 \left(\partial_\mu \phi  \partial^\mu \phi^t ) 
  +
  \frac{1}{2}\cdot \frac{a N}{G} 
  \rho_\mu^2 
  -  i a   \rho^\mu \phi \partial_\mu \phi^t 
  \right)
   \right]
  \, + 
   \frac{1}{2}  {\rm tr} 
  \left[ 
  \left(1-\frac{a}{2}\right) \frac{G}{N} \left(\phi \partial_\mu \phi^t\right)^2
   -\eta\left(\phi \phi^t - \frac{N}{G} \1 \right)
  \right], 
 \label{a-depaction}
 \end{eqnarray}
where ${\rm tr}$ and $\1$ should read ${\rm tr}_{p\times p}$ and $\1_{p\times p}$, respectively,  and $(\rho_\mu)_{ij}=\rho_\mu^a (S^a)_{ij}$ with ${\rm tr}(S^a S^b)=2 \delta^{ab}$, and the $p\times p$ matrix Lagrange multiplier $\eta_{i,j}(x)$ is used for the constraint Eq.(\ref{ONpNp})~\footnote{
  In the broken phase this is simply equivalent to the constraint Eq.(\ref{ONpNp}), while in the unbroken phase 
  the multiplier is only a correct description.   See later discussions.   
  } 
  as in the standard large $N$ arguments of $CP^{N-1}$~\cite{Eichenherr:1978qa,Golo:1978de,DAdda:1978vbw,DAdda:1978dle,Witten:1978bc,Arefeva:1980ms,Haber:1980uy,Kugo:1985jc,Bando:1987br,Weinberg:1997rv,Harada:2003jx} and 
   other Grassmannian models~\cite{Brezin:1980ms,Bando:1996pg}. For $N=4, p=3$ Eq.(\ref{a-depaction}) with $O(4)_{\rm global}\times O(3)_{\rm local}$ is identical to the standard HLS Lagrangian \cite{Bando:1984ej,Bando:1985rf,Bando:1984pw,Fujiwara:1984mp,Bando:1987ym,Bando:1987br,Harada:2003jx,Yamawaki:2016qux} for the rho meson with $[SU(2)_L\times SU(2)_R]_{\rm global}\times [SU(2)_V]_{\rm local}$.
     It is now clear \cite{Yamawaki:2018jvy} that Eq.(\ref{a-depaction}) coincides with that of the conventional ``covariant derivative type'' Lagrangian~\cite{Brezin:1980ms} for a particular choice $a=2$, with    $\phi \phi^t =(N/G) \1$ (see Eq.(\ref{ONpNp})).    
 
 From Eq.(\ref{a-depaction})  the effective potential in the large $N$ limit  for $\langle \hbox{\boldmath$\phi$}_{i,\beta}(x) \rangle=\sqrt{N} v (\delta_{i,j},0)$  (we took $v\ne 0$ real, i.e., the unitary gauge ${\check \rho}(x)=0$) and $\langle \eta_{i,j}(x) \rangle=\eta\, \delta_{i,j}$,  takes the form  (in $D$ dimensions): 
 \begin{eqnarray}
 \frac{V_{\rm eff} \left(v,\eta  \right)}{Np/2}=\eta \left(v^2 -\frac{1}{G} 
 \right) 
  + \int\frac{d^D k}{i (2\pi)^D} \ln \left(k^2-\eta\right),
 \end{eqnarray}
where the ($a$-dependent) 1-PI contributions are sub-leading in the large $N$ limit 
\footnote{This observation is due to  H. Ohki.}
 and therefore the result is {\it independent of the parameter $a$}, in precisely the same form as that of the conventional ``covariant derivative'' parameterization of $CP^{N-1}$ and the Grassmannian models corresponding to $a=2$~\cite{Eichenherr:1978qa,Golo:1978de,DAdda:1978vbw,DAdda:1978dle,Witten:1978bc,Arefeva:1980ms,Haber:1980uy,Kugo:1985jc,Bando:1987br,Weinberg:1997rv,Harada:2003jx,Brezin:1980ms,Bando:1996pg}, and hence 
yields  the  same gap equation: 
 \begin{eqnarray}
\frac{1}{Np}  \frac{\partial V_{\rm eff}}{\partial v}&=& 2 \eta v=0\,,
\label{sta1}\nonumber
\\
\frac{1}{Np} \frac{\partial V_{\rm eff}}{\partial \eta} 
&=& v^2-
\frac{1}{G} 
+ \frac{1}{{G}_{\rm crit}} -v_\eta^2
=0 \,, \label{sta3}
\end{eqnarray}
with (for cutoff $\Lambda$)
~\footnote{
The cutoff $\Lambda$ can be removed for $2\le D <4$ (the theory is renormalizable), introducing the renormalized coupling at renormalization point $\mu$ as 
%\label{renormalized}\\
$1/{G}^{(R)}(\mu)\equiv 1/G-\int \frac{d^Dk}{i(2\pi)^D} \, \frac{1}{\mu^2 - k^2}
\equiv \mu^{D-2}/g^{(R)}(\mu)$, $1/G^{(R)}_{\rm crit} \equiv 
 \int \frac{d^Dk}{i(2\pi)^D} 
 \left(\frac{1}{- k^2}-\frac{1}{\mu^2 - k^2} \right)$
 $= \frac{\Gamma(2- D/2)}{\left(D/2 - 1 \right)}
\cdot \frac{\mu^{D-2}}{(4\pi)^{D/2}}\equiv  \mu^{D-2}/g^{(R)}_{\rm crit} $, s.t., $ 1/G -1/G_{\rm crit} =\mu^{D-2} \left(1/{g}^{(R)}(\mu) -  1/g_{\rm crit}^{(R)} \right)
$ in the gap equation.  The renormalized coupling $ {g}^{(R)}(\mu)$ has an ultraviolet fixed point at $g_{\rm crit}^{(R)}$, $0\le g_{\rm crit}^{(R)}=(4\pi)^{D/2}
\left(D/2 - 1 \right)/\Gamma(2- D/2)  < \infty \, (2\le D<4)$, with the beta function $\beta( {g}^{(R)}(\mu) )=\mu \partial {g}^{(R)}(\mu)/{\partial \mu} =
-(D-2){g}^{(R)}(\mu)  [{g}^{(R)}(\mu) -g_{\rm crit}^{(R)}]/g_{\rm crit}^{(R)}$.  While for $D=4$ the theory is not renormalizable, $1/g_{\rm crit}^{(R)}\sim \Gamma(2-D/2)/(4\pi)^2\Big|_{D \rightarrow 4} \sim \ln (\Lambda^2/\mu^2)/(4\pi)^2$, with
 the remaining log divergence identified in the cutoff notation. 
}
\begin{eqnarray}
 &&\frac{1}{{G}_{\rm crit}}\equiv
\int \frac{d^Dk}{i(2\pi)^D} \, \frac{1}{- k^2}
= \frac{1}{\left(\frac{D}{2} - 1 \right) \Gamma(\frac{D}{2}) }
  \frac{\Lambda^{D-2}}{(4\pi)^{\frac{D}{2}}} , \nonumber
\\
&& v_\eta^2\equiv 
 \int \frac{d^D k}{i (2\pi)^D} \left(\frac{1}{-k^2}- 
 \frac{1}{\eta-k^2} 
 \right)
 =
 \frac{\Gamma(2- \frac{D}{2})}{\frac{D}{2} - 1}
 \cdot 
\frac{\eta^{\frac{D}{2}-1}}{(4\pi)^{\frac{D}{2}}}. \nonumber
\end{eqnarray} 

The gap equation implies as usual the {\it second order phase transition} between two phases, the (weak coupling) phase with the symmetry spontaneously broken which is the same as the classical level, 
and the  (strong coupling) phase with that spontaneously unbroken which is a new phase at quantum level: 
\begin{eqnarray}
&&\mbox{(i)} \quad G < G_{\rm cr} :
 v \neq 0 \ , \ 
 \eta
 =0\,  \left({\rm broken}\,\, {\rm phase}\right)
\nonumber\\
&&
v^2=
\frac{1}{G} - \frac{1}{G_{\rm crit}}
>0 \,\, ,
\label{broken}\\
&&
\mbox{(ii)} \quad G > G_{\rm crit} :
  v = 0 \ , \ 
  \eta
  \ne 0\,  \left({\rm unbroken}\,\, {\rm phase}\right)\nonumber\\
&& 
 v_\eta^2= \frac{1}{G_{\rm crit}} - \frac{1}{G} \, >0\,,
\label{symmetric}
\end{eqnarray}
with the phase transition point $v=\eta=0$. We may define a full decay constant $F_\pi$ at quantum level in the large $N$ limit: 
\begin{eqnarray}
F_\pi^2 \equiv N v^2= N \left(\frac{1}{G} - \frac{1}{G_{\rm crit}}\right) = {f_\pi}^2 -  \frac{N}{\left(\frac{D}{2} - 1 \right) \Gamma(\frac{D}{2}) }
  \frac{\Lambda^{D-2}}{(4\pi)^{\frac{D}{2}}} \quad \rightarrow {f_\pi}^2- N\frac{\Lambda^2}{(4\pi)^2} \,\, \left(D\rightarrow 4\right)\, .
\end{eqnarray}
This indicates that approaching from the broken phase to the critical point, $F_\pi^2 \rightarrow 0$, is due to the power divergence of $1/G_{\rm crit}$(quadratic divergence for $D=4$), similarly to the ``Wilsonian matching'' of the HLS model with the underlying QCD at the UV scale $\Lambda$~\cite{Harada:2003jx}.

\section{Dynamical Generation of Rho Meson}
Now the (amputated) two-point function of $\rho_\mu$ in the large $N$ limit takes the form:
\begin{eqnarray}
&&\Gamma^{(\rho)}_{\mu\nu}(q) = \left(\frac{a}{2}\right)\left(\frac{N}{G}\right) g_{\mu\nu} +  \left(\frac{a}{2}\right)^2 B_{\mu\lambda} (q) \cdot C^\lambda_\nu (q)
,\nonumber\\
&&C_{\mu\nu} (q) = g_{\mu\nu}+ \left(\frac{a}{2} -1\right) \frac{G}{N} B_{\mu\lambda}(q)\cdot   C^\lambda_\nu (q)\,,
\label{twopointa}
\end{eqnarray}
where the four-$\phi$ vertex  $\left(\frac{a}{2} -1\right) \frac{G}{N}$ in our Lagrangian Eq.(\ref{a-depaction}) (second line) gives rise to an infinite sum of the bubble graph contribution $B_{\mu\nu}(q)$; 
\begin{eqnarray}
\frac{1}{N} 
 B_{\mu\nu}(q)
 =
  \frac{1}{2} \int \frac{dk^D}{i (2\pi)^D} 
 \frac{
 (2 k+q)_\mu (2k+q)_\nu
 }{
 \left(k^2-\eta\right)\left((k+q)^2-\eta\right)}
 = q^2 f(q^2,\eta) \cdot \left(
 g_{\mu\nu} - \frac{q_\mu q_\nu}{q^2} 
 \right)  + \left(v^2-\frac{1}{G}\right)  \cdot g_{\mu\nu} \,, 
 \label{bubble} 
 \end{eqnarray} 
 with
 \begin{eqnarray}
f(q^2,\eta)\equiv - 
\frac{\Gamma(2-\frac{D}{2})}
{ 2\left(4\pi\right)^{\frac{D}{2}}\Gamma(2)}
\int_0^1 dx \frac{\left(1- 2x\right)^2}{
\left[x\left(1-x\right) q^2+\eta\right]^{2-\frac{D}{2}
}}\,, \nonumber
\end{eqnarray}
which reads for $D
\rightarrow 4$ ($\epsilon\equiv 2-D/2 \rightarrow 0$ and $1/A^{\epsilon} \simeq 1-\epsilon \ln A$): 
\begin{eqnarray}
f(q^2,0)&=&- \frac{1}{2} \cdot \frac{1}{3 \left(4\pi\right)^2}
\cdot \left[
\ln \left(\frac{\Lambda^2}{q^2}\right) +\frac{8}{3}
\right]\,, \nonumber \\ 
f(0,\eta)&=&- \frac{1}{2} \cdot \frac{1}{3 \left(4\pi\right)^2}\cdot \left[
\ln \left(\frac{\Lambda^2}{\eta}\right)\right] ,
\label{f}
\end{eqnarray}
where we have used the gap equation Eq.(\ref{sta3}) and  identified $\Gamma(\epsilon) \simeq 1/\epsilon \rightarrow \ln \Lambda^2$.\footnote{
The finite part common to both phases are included in the definition of the cutoff $\Lambda$, while the part $+8/3$ is 
an extra one in the broken phase $\eta\equiv 0$, similarly to that in the $CP^{N-1}$ \cite{Weinberg:1997rv}. 
}

\subsection{$a=2$ case}
Note  that for $a=2$, we have $C_{\mu\nu}=g_{\mu\nu}$ in Eq.(\ref{twopointa}), which yields $\Gamma_{\mu\nu}^{(\rho)}(q)$:
\begin{eqnarray}
\frac{\Gamma^{(\rho)}_{\mu\nu}(q)}{N} =\left(\frac{1}{G}\right) g_{\mu\nu} + \frac{B_{\mu\lambda} (q)}{N} \cdot g^\lambda_\nu (q)
=\left(q^2 f(q^2,\eta) +v^2\right) \cdot  \left(
 g_{\mu\nu} - \frac{q_\mu q_\nu}{q^2} 
 \right) + v^2  \cdot \frac{q_\mu q_\nu}{q^2} \,,
 \label{a2Gamma}
\end{eqnarray}
the well-known form of one-loop dominance in the large $N$ limit in the conventional ``covariant derivative'' parameterization for $CP^{N-1}$ model and other Grassmannian models~\cite{Eichenherr:1978qa,Golo:1978de,DAdda:1978vbw,DAdda:1978dle,Witten:1978bc,Arefeva:1980ms,Haber:1980uy,Kugo:1985jc,Bando:1987br,Weinberg:1997rv,Harada:2003jx,Brezin:1980ms,Bando:1996pg}.

For the broken phase $v\ne 0$, $\eta=0$, this is readily inverted to yield the  $\rho_\mu$ propagator for $a=2$:
$\langle \rho_\mu \rho_\nu\rangle(q)\equiv \langle \rho^{ij}_\mu \rho^{ji}_\nu\rangle (q)
= 2 \langle \rho^a_\mu \rho^a_\nu\rangle (q)$:
\begin{eqnarray}
\langle \rho_\mu \rho_\nu
\rangle (q)
 &=&- \Gamma^{(\rho)}_{\mu\nu}(q)^{-1}    =\frac{1}{N}\frac{-f^{-1}(q^2,0) }{q^2  +f^{-1}(q^2,0) v^2}    \left(
 g_{\mu\nu} - \frac{q_\mu q_\nu}{q^2} \right) - \frac{1}{N}\frac{1}{v^2} \frac{q_\mu q_\nu}{q^2}\nonumber \\
 &=&\frac{1}{N} \frac{-f^{-1}(q^2,0) }{q^2  +f^{-1}(q^2,0) v^2}    \left(
 g_{\mu\nu} - \frac{q_\mu q_\nu}{ -f^{-1}(q^2,0) v^2}\right)   =  2 \Delta_{\mu\nu}(q), \nonumber\\
  \Delta_{\mu\nu}(q) &\equiv& 
 \frac{g^2_{_{\rm HLS}}(q^2)}{q^2- M_\rho^2(q^2)}
  \left(
  g_{\mu\nu} - \frac{q_\mu q_\nu}{M_\rho^2(q^2)
  }
  \right),
 \nonumber\\
  M_\rho^2(q^2)\equiv  -f^{-1}(q^2,0) v^2 &=&  g^2_{_{\rm HLS}}(q^2)\cdot 2 F_\pi^2, \quad    g^{-2}_{_{\rm HLS}}(q^2)\equiv  -2N  f(q^2,0)= \frac{N}{3(4\pi)^2} \left[
\ln
\frac{\Lambda^2}{q^2}
+\frac{8}{3}
\right], 
\label{rhopropagatora2}
 \end{eqnarray}
 which is the form of the unitary gauge (we took the unitary gauge $\hat \rho(x)=0$ with $v\ne 0=$ real), with
the physical pole position and  the on-shell HLS coupling given as  $q^2=M_\rho^2= -f^{-1}(M_\rho^2,0) v^2=g^2_{_{\rm HLS}}(M_\rho^2) \cdot 2 F_\pi^2$ and $g^2_{_{\rm HLS}}\equiv g^2_{_{\rm HLS}}(M_\rho^2)$, respectively.
The relation
implies the rho meson mass is generated by the Higgs mechanism 
\begin{eqnarray}
M_\rho^2=g^2_{_{\rm HLS}}\cdot F_\rho^2\,,\quad  F_\rho^2 = 2 \cdot F_\pi^2\,,
\label{Frhoa2}
\end{eqnarray}
where $F_\rho$ is the decay constant  of the would-be NG boson $\hat \rho$ (absorbed into the rho meson in the unitary gauge)  at quantum level, which is to be compared with the tree-level relation Eq.(\ref{frho}), with $a=a^{(p)}=2$. 
The $q^2$ dependence of $M_\rho^2(q^2)$ and   $g^2_{_{\rm HLS}}(q^2)$ may be regarded as
the running mass and the  ({\it asymptotically non-free/infrared free}) running coupling. 
The resultant rho meson mass relation $M_\rho^2=-f^{-1}(M_\rho^2,0) v^2= g^2_{_{\rm HLS}}(M_\rho^2) \cdot 2 F_\pi^2$ is independent of $N$
and can be extrapolated into $N\rightarrow 4$ with $p=3$ for the actual rho meson.

We thus {\it establish the dynamically generation of the rho meson as the HLS gauge boson for $a=2$}~\cite{Yamawaki:2018jvy} in exactly the same way as in the $CP^{N-1}$ model and other Grassmannian models.
 
 In the  unbroken phase,  $v=0, \eta \ne 0$,  on the other hand, $\Gamma^{(\rho)}_{\mu\nu}(q)$ in Eq.(\ref{a2Gamma}) is transverse,  implying the HLS is an {\it  unbroken gauge symmetry}.  Though not invertible as it stands, it is of course inverted by fixing the gauge as usual,  to get  the {\it  massless} propagator $\langle \rho_\mu \rho_\nu\rangle(q)= g^{2}_{_{\rm HLS}}(q^2,\eta)  \cdot \frac{g_{\mu\nu}}{q^2} +$ gauge term, 
 with $g^{-2}_{_{\rm HLS}}(q^2,\eta)\equiv -2N f(q^2,\eta)\simeq -2N f(0,\eta)\equiv g^{-2}_{_{\rm HLS}}(\eta)$ which is analytic at $q^2=0$. 
 $\eta$-dependence may be regarded as the running of the coupling, asymptotically non-free/infrared free, $g^{2}_{_{\rm HLS}}(\eta) \rightarrow 0 \, (\eta \rightarrow 0)$, the same as that in broken phase, see 
 Eq.(\ref{f}).~\footnote{ 
   {\it Without gauge symmetry} ($a=0$), 
 $\langle \alpha_{\mu,||}\, \alpha_{\nu,||}\rangle(q)$ is {\it ill-defined in the unbroken phase} $v=0$, where  the factor $g_{\mu\lambda}+ \frac{G}{N} B_{\mu\lambda}$  is pure transverse  and {\it not invertible},  in accord with the Weinberg-Witten theorem~\cite{Weinberg:1980kq} on the absence of massless  spin $J\geq 1$ particles in the positive definite Hilbert space (no gauge symmetry). 
}  The situation is also the same as the $CP^{N-1}$ and the Grassmannian models.
 Note also that the {\it massless rho meson is stable}, since it  does not decay into the pions which are no longer the NG bosons and have  non-zero mass degenerate with $\check \rho$ (no longer the would-be NG boson)  and other degrees of freedom of $\phi_{i,\beta}$ (corresponding to the 6 constraints in the broken phase, in addition to the 3 $\pi$'s and 3 $\check \rho$'s for $N=4,p=3$), $M_\pi^2=M^2_{\check \rho}=\cdots =\eta\ne 0$. Note that the phase transition is of the second order with $v=\eta=0$ and all the spectra are decoupled (free) massless particles: $M_\rho^2=M^2_{\check \rho}=M_\pi^2=0$ at the phase transition point $G=G_{\rm crit}$ (conformal).

 \subsection{Case for arbitrary value of $a$}
Since we have established the dynamical generation of the rho meson for $a=2$, the next question is wether the conclusion is dependent on the specific value of $a=2$. Here we show that the result is independent of $a$.

For the generic case for arbitrary $a$, the large $N$ dominant diagrams are not just the one-loop but do include  {\it an infinite sum of the bubble diagrams  coming from the extra four-vertex} $\left(\frac{a}{2} -1\right) \frac{G}{N}$ as in Eq.(\ref{twopointa}). $C_{\mu\nu}$ in Eq.(\ref{twopointa})  is solved straightforwardly though tediously (see Ref. \cite{Yamawaki:2018jvy} for details):
From Eqs.(\ref{twopointa}) and (\ref{bubble}) we have
\begin{eqnarray}
 \frac{a}{2}C_{\mu\nu}(q)&=&\frac{a}{2}\left[ g_{\mu\nu} + \left(1- \frac{a}{2}\right) \frac{G}{N} B_{\mu\nu} (q)\right]^{-1} \nonumber\\&=&\left[1 - \left(1-\frac{2}{a}\right) G \left(v^2+ q^2 f \right)\right]^{-1} \left(g_{\mu\nu} -\frac{q_\mu q_\nu}{q^2}\right) 
+ \left[1-\left(1-\frac{2}{a}\right) G v^2\right]^{-1}\frac{q_\mu q_\nu}{q^2} \,, \nonumber\\
\frac{\Gamma^{(\rho)}_{\mu\nu}(q)}{N}&=& \frac{2 }{G} \left(1-\frac{2}{a}\right)^{-1}\left[\frac{a}{2}C_{\mu\nu}- g_{\mu\nu}\right]\,,\nonumber\\
&=& \left[\frac{f^{-1}}{q^2 + v^2 f^{-1} } -\left(1-\frac{2}{a}\right) G 
\right]^{-1} \cdot
\left(g_{\mu\nu} - \frac{q_\mu q_\nu}{q^2}\right) 
+ \left[\frac{1}{v^2}
-\left(1-\frac{2}{a}\right) G \right]^{-1} \cdot 
\frac{q_\mu q_\nu}{q^2},\quad f\equiv f(q^2,\eta).
 \label{Gamma}
\end{eqnarray}
This of course is reduced to Eq.(\ref{a2Gamma}) for $a=2$.

We finally arrive at the {\it dynamically generated propagating HLS gauge boson for any $a$},
 whose propagator in the broken phase, 
  takes the same form of the unitary gauge as that for $a=2$ except for the contact term (to be discussed later):~\cite{Yamawaki:2018jvy} 
   \begin{eqnarray}
\langle \rho_\mu \rho_\nu
\rangle (q)
 &=&- \Gamma^{(\rho)}_{\mu\nu}(q)^{-1}    =
 \left[\frac{- f^{-1}}{q^2 + v^2 f^{-1} } +\left(1-\frac{2}{a}\right) G 
\right] \left(g_{\mu\nu} - \frac{q_\mu q_\nu}{q^2}\right)  
- \left[\frac{1}{v^2}
-\left(1-\frac{2}{a}\right) G \right]\frac{q_\mu q_\nu}{q^2}\nonumber\\
&=& 
   \left(1-\frac{2}{a}\right)\,  \frac{G}{N} \, g_{\mu\nu}+ 2 \Delta_{\mu\nu}(q).
   \label{rhopropagatoruniversal}
  \end{eqnarray}
  which is reduced to Eq.(\ref{rhopropagatora2}) for $a=2$.   
Again the mass relation from $\Delta_{\mu\nu}$ in the last line  of Eq.(\ref{rhopropagatora2}) is independent of $N$ and thus safely extrapolated to the realistic rho meson, $N\rightarrow 4$ with $p=3(=$ fixed). 

Here the physical pole position $q^2=M_\rho^2=
 g^2_{_{\rm HLS}}(M_\rho^2) \cdot 2 F_\pi^2$ and  
 the on-shell HLS coupling $g^2_{_{\rm HLS}}\equiv g^2_{_{\rm HLS}}(M_\rho^2)$ are both {\it independent of $a$};
\begin{eqnarray}
M^2_\rho= g^2_{_{\rm HLS}}\cdot 2 F_\pi^2\,, \quad F_\rho^2=2\cdot F_\pi^2\,,\quad (a-{\rm independent})\, ,
\label{Frho}
\end{eqnarray}
which is the same as Eq.(\ref{Frhoa2}) but now it is  an {\it $a$-independent} result, in contrast to that of the bare quantities at tree level Eq.(\ref{frho}):
$f_\rho^2=a {f_\pi}^2$. This relation is a reminiscence of the KSRF II relation, $M_\rho^2=2 g^2_{\rho\pi\pi} F_\pi^2$, where $g_{\rho\pi\pi}$ is the $\rho\pi\pi$ coupling. In fact in the next section we will  show $g_{\rho\pi\pi}=g_{_{\rm HLS}}$ (``rho-universality'') independently of $a$, and thus derive the KSRF II relation (as well as KSRF I) independently of $a$.

Note that the {\it $a$-dependence is exactly cancelled in the physical part} $\Delta_{\mu\nu}(q)$ as it should be, since $a$ is actually a redundant parameter for the auxiliary field 
$\rho_\mu$. While the $a-$ dependence remains in  the {\it unphysical  contact term}  
$-  \frac{2G}{a N} g_{\mu\nu}$ which corresponds to  the tree $\rho_\mu$ ``propagator'' with tree mass $\frac{aN}{2G}$, it is an artifact in using the auxiliary field $\rho_\mu$ for the composite  field 
$\alpha_{\mu,||}= i\frac{G}{N}\phi\partial_\mu\phi^t $ whose two-point function  is independent of $a$ and  exists even for $a=0$ (without HLS!)  {\it in the broken phase}. (They satisfy an exact relation via Ward-Takahashi identity, $\langle \rho_\mu \rho_\nu\rangle (q) =
\langle  \alpha_{\mu,||}\, \alpha_{\nu,||} \rangle (q)
 -  \frac{2 G}{a N} g_{\mu\nu}
  $~\cite{Yamawaki:2018jvy}).   

Moreover, the whole contact term  is  cancelled in  the $\pi\pi$ scattering.
The $\pi\pi$ scattering amplitude $T_{\mu\nu}(q)$ is  given as $2 T_{\mu\nu}(q) $$=-\frac{G}{N} g_{\mu\nu} $$+ \langle  \alpha_{\mu,||}\, \alpha_{\nu,||} \rangle (q)$, where the first term is from the tree vertex, while  the second term is only from the loop contributions (bubble sum) dominant in the large $N$ limit, $\langle \alpha_{\mu,||}\, \alpha_{\nu,||}\rangle(q)=$
$ \left(i\frac{G}{N}\right)^2 $$\langle \phi\partial_\mu\phi^t 
  \,\, \phi\partial_\nu\phi^t  \rangle (q)$
$=  \left(i\frac{G}{N}\right)^2 $
$\left[B_{\mu\nu}(q) + B_{\mu\lambda}(q)\cdot \left(-\frac{G}{N}\right) B^\lambda_\nu(q) +\cdots\right] $
$= \left(i\frac{G}{N}\right)^2 B_\mu^\lambda(q) $
$\left[g_{\lambda\nu} +\frac{N}{G} \langle \alpha_{\mu,||}\, \alpha_{\nu,||}\rangle(q)
\right]
$
$=\left(g_{\mu\lambda}+ \frac{G}{N} B_{\mu\lambda}\right)^{-1} $$\left(i\frac{G}{N}\right)^2 B^\lambda_\nu(q)$
$= \frac{G}{N} g_{\mu\nu} +2 \Delta_{\mu\nu}(q)$, with $B_{\mu\nu}(q)$ given in Eq.(\ref{bubble}).\footnote{
The four-$\phi$ vertex (both for tree and loop) here is different from  Eq.(\ref{twopointa})
%Gamma) 
for the 
$\rho_\mu$ case: $(\frac{a}{2}-1)\frac{G}{N} +(-\frac{a}{2})\frac{G}{N}=-\frac{G}{N}$ (the additional second term is from the tree rho contribution $(-\frac{ia}{2})(\frac{a}{2}\frac{N}{G})^{-1} (-\frac{ia}{2})$), the same as that for $a=0$ (original nonlinear sigma model without HLS) as it should be independent of the auxiliary field.
}
Then the contact term $\frac{G}{N} g_{\mu\nu}$ is precisely cancelled in $T_{\mu\nu}$, namely 
the {\it VMD for arbitrary value of $a$}. This is compared with the conventional HLS approach where
the VMD for $\pi\pi$ scattering is realized only for $a=4/3$ (not $a=2$!!)~\cite{Harada:2003jx}.

As seen from Eq.(\ref{rhopropagatora2}), the HLS coupling depends on the cutoff $\Lambda$ as it should, since the nonlinear sigma model  is a non-renormalizable model for $D=4$ (see \cite{Weinberg:1997rv} for other formulation).
From 
Eq.(\ref{rhopropagatora2}), with $N=4$ and $q^2=M_\rho^2\simeq (770\, {\rm MeV})^2$, $F_\pi\simeq  92\,{\rm MeV}$, we have $\Lambda=e^{-4/3}\cdot M_\rho \cdot e^{12 \pi^2 F_\pi^2/M_\rho^2}$$ \simeq	 1.1 {\rm GeV} \simeq 4 \pi F_\pi$, roughly the validity scale of the chiral perturbation theory. As an asymptotically non-free theory the kinetic term vanishes $1/g^2_{_{\rm HLS}}(q^2=\mu^2)=-2N f(\mu^2,0) \rightarrow 0$ ($\mu^2\rightarrow {\tilde \Lambda}^2 $) at the Landau pole  $\mu=\tilde \Lambda = e^{4/3} \Lambda
\simeq 4.2$ GeV $\gg \Lambda \gg M_\rho$, where the $\rho_\mu$ returns  to an auxiliary field as a static composite of $\pi$, the situation sometimes referred to as
 ``compositeness condition''~ \cite{Bardeen:1989ds}
 advocated in a reformulation of the top quark condensate model~\cite{Miransky:1988xi}. 
 In this viewpoint the HLS gauge bosons as bound states of $\pi$'s develop the kinetic term as we integrate the higher frequency modes
in the large $N$ limit from $\Lambda^2$ down to the scale $\mu^2$ in the sense of the Wilsonian renormalization group~\cite{Harada:2003jx}. 

  This also implies $g^2_{_{\rm HLS}}(q^2=\mu^2)\rightarrow 0$ ($\mu^2/{\tilde \Lambda}^2 \rightarrow 0$)
 at approaching   the phase transition point $F_\pi^2=N v^2 \rightarrow 0 \, (G \rightarrow G_{\rm cr} -$).  
Thus the rho meson in the broken phase, with $M_\rho$ close enough to the phase transition point, $M_\rho/{\tilde \Lambda}, M_\rho/\Lambda \rightarrow 0$, is to be identified with a gauge boson.~\footnote{ Since the phase transition is second order, the {\it HLS as a gauge symmetry  is crucial not only in the unbroken phase but also in the broken phase near the phase transition point}. (Just on the phase transition point all the
spectra become massless and free, i.e., trivially scale-symmetric.) 
}
The result $g^2_{_{\rm HLS}}\rightarrow 0$ and $M_\rho^2\rightarrow M_\pi^2 (\equiv 0)$ near the phase transition point in the broken phase  is similar to the Vector Manifestation (Ref. \cite{Harada:2003jx} and references cited therein),
{\it both not precisely on the phase transition point} where $\rho$ and $\pi$ are just decoupled massless free particles $M_\rho^2=M_\pi^2=0$. The latter is based on the one-loop ``Wilsonian Matching'' with QCD at $\Lambda$ where the kinetic term is given with the parameter $a=a(\Lambda^2)\simeq 1$, which then runs down as $a (\mu^2) =F_\rho^2(\mu^2)/F_\pi^2(\mu^2)\sim 1 \,(M_\rho^2<\mu^2< \Lambda^2)$ (with $\rho$ loop) and further down to $\pi$ on shell $a(0)=F_\rho^2(M_\rho^2)/F_\pi^2(M_\pi^2=0) =2$ (with the $\rho$ loop decoupled for $F_\pi^2$ in $\mu^2<M_\rho^2$),  in contrast to the present case which is 
 for any $a$ at all orders in the large $N$ limit without $\rho$ loop at all.
 
We then have the effective action 
with kinetic term of rho meson $\rho_\mu$ and/or the composite field $\alpha_{\mu,||}$: 
\begin{eqnarray}
{\cal L}_{\rm kinetic}^{{\rm eff}} =- \frac{1}{4 g^2_{_{\rm HLS}}} \cdot \frac{1}{2} {\rm tr} \rho_{\mu\nu}^2=- \frac{1}{4 g^2_{_{\rm HLS}}} \cdot \frac{1}{2} {\rm tr} \alpha_{\mu\nu,||}^2,
\label{kinetic}
\end{eqnarray}
where $\alpha_{\mu\nu,||}\equiv \partial_\mu\alpha_{\nu,||} - \partial_\nu\alpha_{\mu,||} - i\left[\alpha_{\mu,||}, \alpha_{\nu,||} \right]$,
with $
f^2_\pi=N/G \Rightarrow F^2_\pi=N v^2$.
For $N=4,p=3$ 
Eq.(\ref{kinetic})  is precisely the Skyrme term,
$\frac{1}{32 e^2} {\rm tr}_{_{SU(2)}} [L_\mu, L_\nu]^2$, with $e^2=g^2_{_{\rm HLS}}$, in the $SU(2)$ basis, where we have 
 $\alpha_{\mu\nu,||}= i\left[\alpha_{\mu,\perp}, \alpha_{\nu,\perp} \right]$
and
$L_\mu\equiv \partial_\mu U\cdot U^\dagger$, $U=\xi^2(\pi)=e^{i\pi^a \tau^a/F_\pi}$~\cite{Igarashi:1985et}.

\section{Successful ``$a=2$'' relations realized for any $a$}
 Now we derive all the phenomenologically successful relations for the rho meson independently of $a$.

 The large $N$ Green function for $\rho\pi\pi$ is given as  a  bubble sum, which takes the  
 {\it $a-$independent  form of VMD},  $\langle \rho_\mu(q)\phi(k)\phi(k+q)\rangle$$=\langle  \alpha_{\mu,||}(q)
\, \phi(k)
\, \phi(q+k)
 \rangle$
 $=2 \Delta_{\mu\nu}(q)\cdot (q+2k)^\nu$~\footnote{
  The $\rho\pi\pi$ vertex $\Gamma^{\rho\pi\pi, \nu} (q,k, q+k)\big|^{k^2=(k+q)^2=0}_{\phi-{\rm amputated}}  
 % $$
  =  \frac{a}{2}%$$ 
  \left[
  g_{\mu\nu}
  +
  B_{\mu\lambda}(q)C^\lambda_\nu (q) \cdot  
 \left(\frac{a}{2}-1\right)\frac{G}{N} \right] 
 \cdot (q+2k)^\nu$ is rewritten as $ 
 \left[
  g_{\mu\nu}+ \Gamma^{(\rho)}_{\mu\nu}(q) \left(1- \frac{2}{a}\right)
  \right] 
  \cdot (q+2k)^\nu$, with $\Gamma^{(\rho)}_{\mu\nu}(q)=-\langle \rho_\mu \rho_\nu\rangle(q)^{-1}$.  Then the Green function is
$\langle \rho_\mu\phi\phi\rangle $
 $=\langle \rho_\mu \rho_\nu\rangle(q)$$\cdot  \Gamma^{\rho\pi\pi, \nu} (q,k, q+k)\big|^{k^2=(k+q)^2=0}_{\phi-{\rm amputated}} 
 =\left[\langle \rho_\mu \rho_\nu\rangle(q) - g_{\mu\nu}\cdot \left(1-\frac{2}{a}\right)\frac{G}{N}
\right]
\cdot (q+2k)^\nu 
=2 \Delta_{\mu\nu}(q)\cdot (q+2k)^\nu$, where the $a$-dependence and the contact term are all cancelled out.
\label{directproof}
},
 where the first equality is by the Ward-Takahashi identity
 \footnote{
It follows from the Ward-Takahashi identity
$
0=\int {\cal D} \phi \frac{\delta}{\delta \rho_\mu(x)}$ $ \left(\phi(y) \phi(z)\cdot e^{i S[\phi,\rho_\mu]}\right)
$ $=\int {\cal D}\phi $ $\left(\frac{aN}{2G} \right) $ $\left(\rho_\mu(x)- \alpha_{\mu,||}(x)
\right)\cdot \phi(y) \phi (z)$ $  \cdot e^{i S[\phi,\rho_\mu]},
$
for ${\cal L}$ given in Eq.(\ref{a-depaction}). This shows  $a$-independence of $\langle \rho_\mu \phi\phi\rangle$ even
without explicit calculations in footnote\ref{directproof}, since $\langle \alpha_{\mu,||} \phi\phi\rangle$ is obviously  independent of $a$. 
}.
 We may introduce  ``renormalized'' field $\rho_\mu^{(R)}\equiv$$ g^{-1}_{_{\rm HLS}}(q^2)\cdot \rho_\mu$
by rescaling the ``kinetic term'' to the canonical one, i.e., $\Delta^{(R)}_{\mu\nu}(q)\equiv g^{-2}_{_{\rm HLS}}(q^2) \cdot \Delta_{\mu\nu}(q)$ and  
$  \langle \rho^{(R)}_\mu\phi\phi\rangle$$=\langle  \alpha_{\mu,||}^{(R)}\phi\phi\rangle$$=2 g_{_{\rm HLS}} (q^2)\cdot\Delta^{(R)}_{\mu\nu}(q) \cdot  (q+2k)^\nu $, which is compared with the definition of $g_{\rho\pi\pi}(q^2)$, $\langle \rho^{(R)}_\mu\phi\phi\rangle \equiv 2 g_{\rho\pi\pi}(q^2) \cdot \Delta^{(R)}_{\mu\nu}(q) \cdot (q+2k)^\nu$, 
resulting in the $\rho$ universality  {\it independently of $a$}:
\begin{eqnarray}
g_{\rho\pi\pi}(q^2)=g_{_{\rm HLS}} (q^2) \quad \left( \rho\,\, {\rm universality}\right).
\label{universality}
\end{eqnarray} 

It then leads to the KSRF relations (generalized for $\forall q^2$) {\it independently of $a$}
\footnote{
This is consistent with the fact that  the KSRF I relation 
is a low energy theorem of the HLS valid  for any $a$~\cite{Bando:1984pw,Bando:1987br},
 which is proved to all order of loop expansion~\cite{Harada:1993jk}. 
}:
\begin{eqnarray}
g_\rho(q^2)&=& M_\rho(q^2) F_\rho=2 g_{\rho\pi\pi}(q^2) F_\pi^2\,\, \left({\rm KSRF \,I} \right),\\
M_\rho^2(q^2)&=& 2 g_{\rho\pi\pi}^2(q^2) F_\pi^2\,\, \left({\rm KSRF \,II} \right),
\end{eqnarray}
with $\langle 0|J^{{\rm em}}_\mu |\rho^{(R)} (q^2)\rangle\equiv g_\rho(q^2)$$ \epsilon_\mu(q)=M_\rho(q^2) F_\rho\epsilon_\mu(q)$.

The VMD for the electromagnetic form factor $F_{_{{\cal B}\pi\pi}}(q^2)$ follows also $a-$independently, similarly to the VMD in the $\pi\pi$ scattering. Here the photon field $ {\cal B}_\mu$ is introduced by gauging  $H_{\rm global}$, $D_\mu \phi \Rightarrow   \partial_\mu \phi -i\rho_\mu \phi+ i \phi {\cal B}_\mu$. It has  contributions from the $ {\cal B}_\mu-\rho_\mu$ mixing  and  from the ``direct coupling'' to $\alpha_{\mu,||}$ (with the tree contact term cancelled by the bubble sum as in the $\pi\pi$ scattering),
 both coupled to the identical VMD Green functions 
$\langle  \rho_\mu^{(R)}\phi\phi\rangle=\langle  \alpha_{\mu,||}^{(R)}\phi\phi\rangle$ in a linear combination to cancel the 
$a$ dependence\footnote{
 $2 F_{_{{\cal B}\pi\pi}}(q^2)$$ \left(q+2k\right)_\mu=%$$
\langle J^{{\rm em}}_\mu(q)\, \phi(k)\,\phi(k+q)\rangle\big|^{k^2=(k+q)^2=0}
_{\phi-{\rm amputated}} 
=- g_\rho(q^2)$ $ \left[
\frac{a}{2}
\langle \rho^{(R)}_\mu
\, \phi\, \phi
\rangle
 +\left(1-\frac{a}{2}\right) \langle  \alpha_{\mu,||}^{(R)}\, \phi\, \phi \rangle\right]
\left(q+2k\right)^\nu
=2 \cdot \left(-M_\rho^2(q^2)\cdot \Delta^{(R)}_{\mu\nu}(q)\right)\cdot  (q+2k)^\nu
$. 
}:
\begin{eqnarray}
F_{_{{\cal B}\pi\pi}}(q^2) =\frac{M_\rho^2(q^2)}{M_\rho^2(q^2)-q^2},\quad F_{_{{\cal B}\pi\pi}}(0)= 1.
\label{formfactor}
\end{eqnarray}
Thus the {\it VMD is realized, independently of $a$}.~\footnote{
 Although  it takes the same form as the naive VMD, $F_{_{{\cal B}\pi\pi}}(q^2) \approx 
 \frac{M_\rho^2}{M_\rho^2-q^2}$ near the on-shell,  $ q^2\approx M_\rho^2$, $M_\rho(q^2)$ here has   log $q^2$ dependence as in Eq.(\ref{rhopropagatora2}). 
 % rhopropagatoruniversal}). 
 Actually, such a $q^2$ dependence is  necessary for the modern version of the  
VMD in both the space-like and the time-like momentum regions, see e.g.,~\cite{Gounaris:1968mw,Harada:1995sj,Benayoun:2011mm}.
}

\section{Conclusion and Discussions}
To conclude we have proved that the rho meson is a dynamical gauge boson of the HLS  $O(3)_{\rm local}\simeq  [SU(2)_V]_{\rm local}$ by the large $N$ 
dynamics of the  model $G/H=O(N)/[O(N-3)\times O(3)]$,   
with all the successful ``$a=2$ results'' being realized purely dynamically {\it independently  of $N$} for {\it any value of $a$},
thus  safely extrapolated to $N=4$, 
 $O(4)/O(3)\simeq O(4)_{\rm global} \times O(3)_{\rm local}  \simeq  [SU(2)_L\times SU(2)_R]_{\rm global}\times [SU(2)_V]_{\rm local}$.
 
 The ``$a=2$ results''  originally obtained for particular choice of $a=2$~\cite{Bando:1984ej,Bando:1985rf,
Bando:1984pw,Fujiwara:1984mp,Bando:1987ym}  are now 
clear to be artifacts of the combined use of the $a-$dependent  {\it tree-level} rho meson mass term 
 and the {\it ad hoc added kinetic term}  which was {\it assumed} to be generated at quantum level 
 {\it without affecting the pole structure} of the dynamically generated propagator.
Actually, as we demonstrated, the tree-level parameter is no longer the true one of the pole  at quantum level when  the kinetic term is generated, namely the pole position (and residue as well) of the full propagator is shifted from
the tree level one in such a way that the $a-$ dependence is totally cancelled out. 
Actually, the parameter {\it $a$ is a redundant parameter for the auxiliary field} $\rho_\mu$ and is irrelevant to the physical results at quantum level as it should be for the correct calculations.
The results of the present paper revealed that it is indeed the case in the large $N$ limit.

Further  implications of the results are:  Once the rho kinetic term is generated, Eq.(\ref{kinetic}), it stabilizes the Skyrmion without ad hoc Skyrme term and hence the nonlinear sigma model in the large $N$ limit perfectly describes via HLS the low energy QCD for 
$\pi,\rho, N$ at the scale $\lesssim \Lambda\simeq 4 \pi f_\pi$ without explicit recourse to the QCD. 

The dynamically generated kinetic term, with the induced gauge coupling $g^2_{\rm HLS}(q^2)$ being asymptotically non-free/infrared free in both broken and unbroken phases, has a cutoff $\Lambda \simeq 4\pi f_\pi \gg M_\rho$
(and Landau pole $\tilde \Lambda$), so that the rho meson
is sitting {\it near the second order phase transition point} 
 as a  composite HLS {\it gauge boson}  to  be matched with the underlying QCD.  This implies~\cite{Yamawaki:2018jvy}  that   the large $N$ dynamics reveals  the HLS as a  ``magnetic gauge theory'' (infrared free in both phases) dual to the underlying QCD as the    ``electric gauge theory''~\cite{Harada:1999zj,Harada:2003jx,Komargodski:2010mc,Kitano:2011zk}, similarly to the Seiberg duality in the SUSY QCD~\cite{Seiberg:1994bz}. 

If  the HLS as the unbroken magnetic gauge theory  is realized, say in hot/dense QCD, we would have a new possibility for the 
chiral symmetry restored hadronic phase having massless
rho meson and massive $\pi, \check \rho$~\cite{Yamawaki:2018jvy}, which is contrasted with $M_\rho^2\rightarrow M^2_\pi (\equiv 0)$ near the phase transition point in the broken phase 
(not precisely on the phase transition point) similarly to the ``Vector Manifestation'' as described in the text.

 It was frequently emphasized that the large $N$ results
are valid even for the small $N$ at least qualitatively as mentioned in the footnote \ref{largeN}.  The result of the present paper is a yet another proof of this statement, and even
more, quantitatively not just qualitatively, in perfect agreement with the experimental facts of the rho meson.

This  further implies the dynamical HLS bosons in other system described by the large $N$ Grassmannian models. A notable case of such is
the Standard Model (SM) Higgs Lagrangian, re-parameterized \cite{Fukano:2015zua}
 as a scale-invariant version of the model $G/H=O(4)/(3)\simeq O(4)_{\rm global} \times O(3)_{\rm local}$, is precisely the same as the rho meson   
case, except for an extra mode, pseudo-dilaton (SM Higgs boson) to make the model (approximately) scale-invariant (Having no indices running through $N$, it is  irrelevant to the SM rho physics in the large $N$ limit)~\cite{Yamawaki:2018jvy}. This justifies
the basic assumption~\cite{Matsuzaki:2016iyq} that there exists a rho meson-like vector boson within the SM (``SM rho'')  which  stabilizes 
 a skyrmion (``SM skyrmion'') as a candidate  for the dark matter existing even within the SM.

\acknowledgments 
We would like to thank T. Kugo who made invaluable help for 
the preliminary results in Ref.\cite{Yamawaki:2018jvy}.
Thanks also go to H. Ohki
for valuable discussions and comments. Special thanks go to Mannque Rho for useful questions and inviting the contribution  to the special issue of 
 the ``Symmetry''.


\begin{thebibliography}{99} 
   

%\cite{Bando:1984ej}
\bibitem{Bando:1984ej} 
 M.~Bando, T.~Kugo, S.~Uehara, K.~Yamawaki, and T.~Yanagida, 
%{\it Is rho meson a dynamical gauge boson of hidden local symmetry?},  
%{\em 
Phys. Rev. Lett. 
%} 
 {\bf 54} (1985) 1215.
 
 
   
 
\bibitem{Bando:1985rf}
M.~Bando, T.~Kugo, and K.~Yamawaki, 
%{\it On the vector mesons as dynamical gauge bosons of hidden local symmetries},  
%{\em 
Nucl. Phys. %} 
 {\bf B259}  (1985) 493.
 
 %\cite{Bando:1984pw}
\bibitem{Bando:1984pw} 
  M.~Bando, T.~Kugo and K.~Yamawaki,
  %``Composite Gauge Bosons and 'Low-energy Theorems' of Hidden Local Symmetries,''
  Prog.\ Theor.\ Phys.\  {\bf 73}, 1541 (1985).
  %doi:10.1143/PTP.73.1541
  %%CITATION = doi:10.1143/PTP.73.1541;%% 

%\cite{Fujiwara:1984mp}
\bibitem{Fujiwara:1984mp} 
  T.~Fujiwara, T.~Kugo, H.~Terao, S.~Uehara and K.~Yamawaki,
  %``Nonabelian Anomaly and Vector Mesons as Dynamical Gauge Bosons of Hidden Local Symmetries,''
  Prog.\ Theor.\ Phys.\  {\bf 73}, 926 (1985);
 % doi:10.1143/PTP.73.926
  %%CITATION = doi:10.1143/PTP.73.926;%%

 %\cite{Bando:1987ym}
\bibitem{Bando:1987ym} 
  M.~Bando, T.~Fujiwara and K.~Yamawaki,
  %``Generalized Hidden Local Symmetry and the A1 Meson,''
  Prog.\ Theor.\ Phys.\  {\bf 79}, 1140 (1988).
%  doi:10.1143/PTP.79.1140
  %%CITATION = doi:10.1143/PTP.79.1140;%% 
  
   
  
  
%\cite{Bando:1987br}
\bibitem{Bando:1987br} 
  M.~Bando, T.~Kugo and K.~Yamawaki,
  %``Nonlinear Realization and Hidden Local Symmetries,''
  Phys.\ Rept.\  {\bf 164}, 217 (1988).
  %doi:10.1016/0370-1573(88)90019-1
  %%CITATION = doi:10.1016/0370-1573(88)90019-1;%%

  
 %\cite{Harada:2003jx}
 \bibitem{Harada:2003jx} 
  M.~Harada and K.~Yamawaki,
  %``Hidden local symmetry at loop: A New perspective of composite gauge boson and chiral phase transition,''
  Phys.\ Rept.\  {\bf 381}, 1 (2003).
 % doi:10.1016/S0370-1573(03)00139-X
 % [hep-ph/0302103].
  %%CITATION = doi:10.1016/S0370-1573(03)00139-X;%% 
  
    %\cite{Yamawaki:2016qux}
\bibitem{Yamawaki:2016qux} 
  K.~Yamawaki,
  %``Hidden Local Symmetry and Beyond,''
  Int.\ J.\ Mod.\ Phys.\ E {\bf 26}, no. 01n02, 1740032 (2017). %, and references cited therein.
 % doi:10.1142/S0218301317400328
%  [arXiv:1609.03715 [hep-ph]].
  %%CITATION = doi:10.1142/S0218301317400328;%%
  
  %\cite{Coleman:1969sm}
\bibitem{Coleman:1969sm} 
  S.~R.~Coleman, J.~Wess and B.~Zumino,
  %``Structure of phenomenological Lagrangians. 1.,''
  Phys.\ Rev.\  {\bf 177}, 2239 (1969).
 % doi:10.1103/PhysRev.177.2239
  %%CITATION = doi:10.1103/PhysRev.177.2239;%%
  
 %\cite{Callan:1969sn}
\bibitem{Callan:1969sn} 
  C.~G.~Callan, Jr., S.~R.~Coleman, J.~Wess and B.~Zumino,
  %``Structure of phenomenological Lagrangians. 2.,''
  Phys.\ Rev.\  {\bf 177}, 2247 (1969).
 % doi:10.1103/PhysRev.177.2247
  %%CITATION = doi:10.1103/PhysRev.177.2247;%%
  
 
   %\cite{Yamawaki:2018jvy}
\bibitem{Yamawaki:2018jvy} 
  K.~Yamawaki,
  %``Dynamical Gauge Boson of Hidden Local Symmetry within the Standard Model,''
  arXiv:1803.07271 [hep-ph].
  %%CITATION = ARXIV:1803.07271;%%
 
   
%\cite{Eichenherr:1978qa}
\bibitem{Eichenherr:1978qa} 
  H.~Eichenherr,
%  ``SU(N) Invariant Nonlinear Sigma Models,''
  Nucl.\ Phys.\ B {\bf 146}, 215 (1978)
  Erratum: [%Nucl.\ Phys.\ 
  B {\bf 155}, 544 (1979)]. 
%  doi:10.1016/0550-3213(79)90287-6, 10.1016/0550-3213(78)90439-X
  %%CITATION = doi:10.1016/0550-3213(79)90287-6, 10.1016/0550-3213(78)90439-X;%%
%

%\cite{Golo:1978de}
\bibitem{Golo:1978de} 
  V.~L.~Golo and A.~M.~Perelomov,
%  ``Solution of the Duality Equations for the Two-Dimensional SU(N) Invariant Chiral Model,''
  Phys.\ Lett.\  {\bf 79B}, 112 (1978). 
 % doi:10.1016/0370-2693(78)90447-1
  %%CITATION = doi:10.1016/0370-2693(78)90447-1;%%

%\cite{DAdda:1978vbw}    
\bibitem{DAdda:1978vbw} 
  A.~D'Adda, M.~Luscher and P.~Di Vecchia,
%  ``A 1/n Expandable Series of Nonlinear Sigma Models with Instantons,''
  Nucl.\ Phys.\ B {\bf 146}, 63 (1978). 
 % doi:10.1016/0550-3213(78)90432-7
  %%CITATION = doi:10.1016/0550-3213(78)90432-7;%%
%


%\cite{DAdda:1978dle}
\bibitem{DAdda:1978dle} 
  A.~D'Adda, P.~Di Vecchia and M.~Luscher,
%  ``Confinement and Chiral Symmetry Breaking in CP**n-1 Models with Quarks,''
  Nucl.\ Phys.\ B {\bf 152}, 125 (1979). 
  %doi:10.1016/0550-3213(79)90083-X
  %%CITATION = doi:10.1016/0550-3213(79)90083-X;%%
%

%\cite{Witten:1978bc}
\bibitem{Witten:1978bc} 
  E.~Witten,
 % ``Instantons, the Quark Model, and the 1/n Expansion,''
  Nucl.\ Phys.\ B {\bf 149}, 285 (1979). 
 % doi:10.1016/0550-3213(79)90243-8
  %%CITATION = doi:10.1016/0550-3213(79)90243-8;%%  
%  

 %\cite{Arefeva:1980ms}
\bibitem{Arefeva:1980ms} 
  I.~Y.~Arefeva and S.~I.~Azakov,
%  ``Renormalization And Phase Transition In The Quantum Cp**(n-1) Model (d = 2, 3),''
  Nucl.\ Phys.\ B {\bf 162}, 298 (1980).
  %doi:10.1016/0550-3213(80)90266-7
  %%CITATION = doi:10.1016/0550-3213(80)90266-7;%%
% 

%\cite{Haber:1980uy}
\bibitem{Haber:1980uy} 
  H.~E.~Haber, I.~Hinchliffe and E.~Rabinovici,
  %``The {CP}**(n-1) Model With Unconstrained Variables,''
  Nucl.\ Phys.\ B {\bf 172}, 458 (1980).
 % doi:10.1016/0550-3213(80)90178-9
  %%CITATION = doi:10.1016/0550-3213(80)90178-9;%%
  
  %\cite{Kugo:1985jc}
\bibitem{Kugo:1985jc} 
  T.~Kugo, H.~Terao and S.~Uehara,
  %``Dynamical Gauge Bosons And Hidden Local Symmetries,''
  Prog.\ Theor.\ Phys.\ Suppl.\  {\bf 85}, 122 (1985).
  
%\cite{Weinberg:1997rv}
\bibitem{Weinberg:1997rv} 
  S.~Weinberg,
%  ``Effective field theories in the large N limit,''
  Phys.\ Rev.\ D {\bf 56}, 2303 (1997).
  %doi:10.1103/PhysRevD.56.2303
 % [hep-th/9706042].
  %%CITATION = doi:10.1103/PhysRevD.56.2303;%%
  
    
%\cite{Brezin:1980ms}
\bibitem{Brezin:1980ms} 
  E.~Brezin, S.~Hikami and J.~Zinn-Justin,
  %``Generalized Nonlinear $\Sigma$ Models With Gauge Invariance,''
  Nucl.\ Phys.\ B {\bf 165}, 528 (1980).
 % doi:10.1016/0550-3213(80)90047-4
  %%CITATION = doi:10.1016/0550-3213(80)90047-4;%%

        
  %\cite{Bando:1996pg}
\bibitem{Bando:1996pg} 
  M.~Bando, Y.~Taniguchi and S.~Tanimura,
  %``Dynamical gauge boson and strong - weak reciprocity,''
  Prog.\ Theor.\ Phys.\  {\bf 97}, 665 (1997).
%  doi:10.1143/PTP.97.665
%  [hep-th/9610244].
  %%CITATION = doi:10.1143/PTP.97.665;%%       
  
 
 % \bibitem{startingLag} 
 %This Lagrangian was studied extensively in \cite{Yamawaki:2018jvy}  though mainly in a different context of physics.

 %\cite{Bardeen:1989ds}
\bibitem{Bardeen:1989ds} 
  W.~A.~Bardeen, C.~T.~Hill and M.~Lindner,
  %``Minimal Dynamical Symmetry Breaking of the Standard Model,''
  Phys.\ Rev.\ D {\bf 41}, 1647 (1990).
  %doi:10.1103/PhysRevD.41.1647
  %%CITATION = doi:10.1103/PhysRevD.41.1647;%%    
 
 %\cite{Miransky:1988xi}
\bibitem{Miransky:1988xi} 
  V.~A.~Miransky, M.~Tanabashi and K.~Yamawaki,
 % ``Dynamical Electroweak Symmetry Breaking with Large Anomalous Dimension and t Quark Condensate,''
  Phys.\ Lett.\ B {\bf 221}, 177 (1989);
  % doi:10.1016/0370-2693(89)91494-9
    %\cite{Miransky:1989ds}
%\bibitem{Miransky:1989ds} 
 % V.~A.~Miransky, M.~Tanabashi and K.~Yamawaki,
 % ``Is the t Quark Responsible for the Mass of W and Z Bosons?,''
  Mod.\ Phys.\ Lett.\ A {\bf 4}, 1043 (1989).
  %doi:10.1142/S0217732389001210
 
 
    
%\cite{Weinberg:1980kq}
\bibitem{Weinberg:1980kq} 
  S.~Weinberg and E.~Witten,
  %``Limits on Massless Particles,''
  Phys.\ Lett.\  {\bf 96B}, 59 (1980).
  %doi:10.1016/0370-2693(80)90212-9
  %%CITATION = doi:10.1016/0370-2693(80)90212-9;%%


 
%\cite{Igarashi:1985et}
\bibitem{Igarashi:1985et} 
  Y.~Igarashi, M.~Johmura, A.~Kobayashi, H.~Otsu, T.~Sato and S.~Sawada,
  %``Stabilization of Skyrmions via $\rho$ Mesons,''
  Nucl.\ Phys.\ B {\bf 259}, 721 (1985).
 % doi:10.1016/0550-3213(85)90010-0
  %%CITATION = doi:10.1016/0550-3213(85)90010-0;%% 
   
 
   %\cite{Harada:1993jk}
\bibitem{Harada:1993jk} 
  M.~Harada, T.~Kugo and K.~Yamawaki,
  %``Proving the low-energy theorem of hidden local symmetry,''
  Phys.\ Rev.\ Lett.\  {\bf 71}, 1299 (1993);
  %doi:10.1103/PhysRevLett.71.1299
 % [hep-ph/9303257].
  %%CITATION = doi:10.1103/PhysRevLett.71.1299;%%
  %\cite{Harada:1993qi}
%\bibitem{Harada:1993qi} 
%  M.~Harada, T.~Kugo and K.~Yamawaki,
  %``Low-energy theorems of hidden local symmetries,''
  Prog.\ Theor.\ Phys.\  {\bf 91}, 801 (1994)
 % doi:10.1143/PTP.91.801, 10.1143/ptp/91.4.801
 % [hep-ph/9303258].
  %%CITATION = doi:10.1143/PTP.91.801, 10.1143/ptp/91.4.801;%%    
  

  

  
  %\cite{Gounaris:1968mw}
\bibitem{Gounaris:1968mw} 
  G.~J.~Gounaris and J.~J.~Sakurai,
  %``Finite width corrections to the vector meson dominance prediction for rho ---> e+ e-,''
  Phys.\ Rev.\ Lett.\  {\bf 21}, 244 (1968).
%  doi:10.1103/PhysRevLett.21.244
  %%CITATION = doi:10.1103/PhysRevLett.21.244;%%
  
  %\cite{Harada:1995sj}
\bibitem{Harada:1995sj} 
  M.~Harada and J.~Schechter,
  %``Effects of symmetry breaking on the strong and electroweak interactions of the vector nonet,''
  Phys.\ Rev.\ D {\bf 54}, 3394 (1996).
 % doi:10.1103/PhysRevD.54.3394
  %[hep-ph/9506473].
  %%CITATION = doi:10.1103/PhysRevD.54.3394;%%
  
%\cite{Benayoun:2011mm}
\bibitem{Benayoun:2011mm} 
  M.~Benayoun, P.~David, L.~DelBuono and F.~Jegerlehner,
  %``Upgraded Breaking Of The HLS Model: A Full Solution to the $\tau^-e^+e^-$ and $\phi$ Decay Issues And Its Consequences On g-2 VMD Estimates,''
  Eur.\ Phys.\ J.\ C {\bf 72}, 1848 (2012).
%  doi:10.1140/epjc/s10052-011-1848-2
 % [arXiv:1106.1315 [hep-ph]].
  %%CITATION = doi:10.1140/epjc/s10052-011-1848-2;%%  

  
 %\cite{Harada:1999zj}
\bibitem{Harada:1999zj} 
  M.~Harada and K.~Yamawaki,
  %``Conformal phase transition and fate of the hidden local symmetry in large N(f) QCD,''
  Phys.\ Rev.\ Lett.\  {\bf 83}, 3374 (1999).
 % doi:10.1103/PhysRevLett.83.3374.
 % [hep-ph/9906445].
  %%CITATION = doi:10.1103/PhysRevLett.83.3374;%% 
  
  %\cite{Komargodski:2010mc}
\bibitem{Komargodski:2010mc} 
  Z.~Komargodski,
  %``Vector Mesons and an Interpretation of Seiberg Duality,''
  JHEP {\bf 1102}, 019 (2011).
 % doi:10.1007/JHEP02(2011)019
  %[arXiv:1010.4105 [hep-th]].
  %%CITATION = doi:10.1007/JHEP02(2011)019;%%
  
 %\cite{Kitano:2011zk}
\bibitem{Kitano:2011zk} 
  R.~Kitano,
  %``Hidden local symmetry and color confinement,''
  JHEP {\bf 1111}, 124 (2011).
 % doi:10.1007/JHEP11(2011)124
 % [arXiv:1109.6158 [hep-th]].
  %%CITATION = doi:10.1007/JHEP11(2011)124;%% 
  
  
%\cite{Seiberg:1994bz}
\bibitem{Seiberg:1994bz} 
  N.~Seiberg,
  %``Exact results on the space of vacua of four-dimensional SUSY gauge theories,''
  Phys.\ Rev.\ D {\bf 49}, 6857 (1994).
 % doi:10.1103/PhysRevD.49.6857
%  [hep-th/9402044].
  %%CITATION = doi:10.1103/PhysRevD.49.6857;%%   
  
  %\cite{Fukano:2015zua}
\bibitem{Fukano:2015zua} 
  H.~S.~Fukano, S.~Matsuzaki, K.~Terashi and K.~Yamawaki,
  %``Conformal Barrier and Hidden Local Symmetry Constraints: Walking Technirhos in LHC Diboson Channels,''
  Nucl.\ Phys.\ B {\bf 904}, 400 (2016).
 % doi:10.1016/j.nuclphysb.2016.01.020
  %[arXiv:1510.08184 [hep-ph]].
  %%CITATION = doi:10.1016/j.nuclphysb.2016.01.020;%%
  
%\cite{Matsuzaki:2016iyq}
\bibitem{Matsuzaki:2016iyq} 
  S.~Matsuzaki, H.~Ohki and K.~Yamawaki,
  %``Dark Side of the Standard Model: Dormant New Physics Awaken,''
  arXiv:1608.03691 [hep-ph].
  %%CITATION = ARXIV:1608.03691;%%   
   
  
    
\end{thebibliography}
\end{document}